\documentclass[a4paper,11pt]{article}
\usepackage{pos}

\title{Large neutrino telescope Baikal-GVD: recent status}
\ShortTitle{Baikal-GVD status 2023}

\author[a]{V.M.~Aynutdinov}
\author[b]{V.A.~Allakhverdyan}
\author[a]{A.D.~Avrorin}
\author[a]{A.V.~Avrorin}
\author[c,d]{Z.~Barda\v{c}ov\'{a}}
\author[b]{I.A.~Belolaptikov}
\author[a]{E.A.~Bondarev}
\author[b]{I.V.~Borina}
\author[e]{N.M.~Budnev}
\author[l]{V.A.~Chadymov}
\author[f]{A.S.~Chepurnov}
\author[b,g]{V.Y.~Dik}
\author[a]{G.V.~Domogatsky}
\author[a]{A.A.~Doroshenko}
\author*[c]{R.~Dvornick\'{y}}
\author[e]{A.N.~Dyachok}
\author[a]{Zh.-A.M.~Dzhilkibaev}
\author[c,d]{E.~Eckerov\'{a}}
\author[b]{T.V.~Elzhov}
\author[d]{L.~Fajt}
\author[l]{V.N. Fomin}
\author[e]{A.R.~Gafarov}
\author[a]{K.V.~Golubkov}
\author[b]{N.S.~Gorshkov}
\author[e]{T.I.~Gress}
\author[h]{K.G.~Kebkal}
\author[a]{I.V.~Kharuk}
\author[b]{E.V.~Khramov}
\author[b]{M.M.~Kolbin}
\author[i]{S.O.~Koligaev}
\author[b]{K.V.~Konischev}
\author[b]{A.V.~Korobchenko}
\author[a]{A.P.~Koshechkin}
\author[f]{V.A.~Kozhin}
\author[b]{M.V.~Kruglov}
\author[j]{V.F.~Kulepov}
\author[e]{Y.E.~Lemeshev}
\author[a,\dagger]{M.B.~Milenin}
\author[e]{R.R.~Mirgazov}
\author[b]{D.V.~Naumov}
\author[f]{A.S.~Nikolaev}
\author[a]{D.P.~Petukhov}
\author[b]{E.N.~Pliskovsky}
\author[k]{M.I.~Rozanov}
\author[e]{E.V.~Ryabov}
\author[a]{G.B.~Safronov}
\author[b,g]{D.~Seitova}
\author[b]{B.A.~Shaybonov}
\author[a]{M.D.~Shelepov}
\author[a]{S.D.~Shilkin}
\author[f]{E.V.~Shirokov}
\author[c,d]{F.~\v{S}imkovic}
\author[b]{A.E.~Sirenko}
\author[f]{A.V.~Skurikhin}
\author[b]{A.G.~Solovjev}
\author[b]{M.N.~Sorokovikov}
\author[d]{I.~\v{S}tekl}
\author[a]{A.P.~Stromakov}
\author[a]{O.V.~Suvorova}
\author[e]{V.A.~Tabolenko}
\author[b]{B.B.~Ulzutuev}
\author[b]{Y.V.~Yablokova}
\author[a]{D.N.~Zaborov}
\author[b]{S.I.~Zavyalov}
\author[b]{D.Y.~Zvezdov}

\affiliation[a]{Institute for Nuclear Research, Russian Academy of Sciences, Moscow, 117312, Russia}
\affiliation[b]{Joint Institute for Nuclear Research, Dubna, 141980, Russia}
\affiliation[c]{Comenius University, 81499 Bratislava, Slovakia}
\affiliation[d]{Czech Technical University in Prague, Institute of Experimental and Applied Physics, 11000 Prague, Czech Republic}
\affiliation[e]{Irkutsk State University, Irkutsk, 664003, Russia}
\affiliation[f]{Skobeltsyn Institute of Nuclear Physics, Moscow State University, Moscow, 119991, Russia}
\affiliation[g]{Institute of Nuclear Physics of the Ministry of Energy of the Republic of Kazakhstan, Almaty, 050032, Kazakhstan}
\affiliation[h]{LATENA, St. Petersburg, 199106, Russia}
\affiliation[i]{INFRAD, Dubna, 141981, Russia}
\affiliation[j]{Nizhny Novgorod State Technical University, Nizhny Novgorod, 603950, Russia}
\affiliation[k]{St.~Petersburg State Marine Technical University, St.~Petersburg, 190008, Russia}
\affiliation[l]{Moscow, free researcher}

\note[\textdagger]{Deceased.}

\emailAdd{dvornicky@fmph.uniba.sk}

\abstract{The Baikal-GVD is a deep-underwater neutrino telescope being constructed in Lake Baikal. After the winter 2023 deployment campaign the detector consists of 3456 optical modules installed on 96 vertical strings. The status of the detector and progress in data analysis are discussed in present report. The Baikal-GVD data collected in 2018-2022 indicate the presence of cosmic neutrino flux in high-energy cascade events consistent with observations by the IceCube neutrino telescope. Analysis of track-like events results in identification of first high-energy muon neutrino candidates. These and other results from 2018-2022 data samples are reviewed in this report.}

\ConferenceLogo{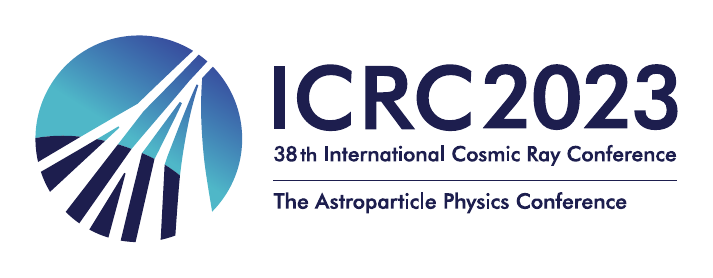}

\FullConference{%
38th International Cosmic Ray Conference (ICRC2023)\\
  26 July - 3 August, 2023\\
  Nagoya, Japan}

\begin{document}
\maketitle

\section{Introduction}

Detection of high-energy neutrinos of astrophysical origin has emerged recently thanks to the construction of gigaton volume neutrino telescopes deep in ice and under water at both the Southern (IceCube) and Northern (Baikal-GVD, ANTARES, and KM3NeT) Hemispheres. The aim of high-energy neutrino astronomy is to reveal the astrophysical sources of the most energetic particles in the Universe which remain unknown yet. Neutrinos, unlike cosmic rays and $\gamma$-rays are not influenced by either the intergalactic magnetic fields or the cosmic microwave background radiation, which makes them perfect for identification of distant high-energy particle sources.   

The IceCube experiment \cite{IceCube2013} has shown the existence of neutrinos of cosmic origin by means of the detection of a diffuse flux exceeding the expected background. Recently first associations of sources established with radio, optical, x-ray and gamma-astronomy observations to high-energy neutrino have been obtained \cite{IceCube2018, IceCubeNGC1068}. Indeed the multi-messenger astronomy turns out to be a very promising and dynamic research field in search of new neutrino sources of astrophysical nature.

In this contribution a progress on the construction and performance of the deep water neutrino telescope Baikal-GVD is reported. Recent developments in track-like and cascade-like event analysis are reviewed. Results on the diffuse neutrino flux in cascade channel are discussed.  

\section{Baikal-GVD detector status}

A cubic kilometer neutrino telescope Baikal-GVD is under construction in the southern part of Lake Baikal since 2016 \cite{BaikalGVD}. This facility is located about 3.6 km offshore at 51$^\circ$ 46'N and 104$^\circ$ 24'E coordinates. The lakebed reaches its plateu at a nearly constant depth of 1\,366 m. The Baikal-GVD telescope is a 3-dimensional array of photo-sensitive elements - optical modules (OMs). The goal is to detect Cherenkov light from secondary muons and cascades produced in neutrino interactions in the water of the lake.  The detector layout is  optimized for the measurement of astrophysical neutrinos in the TeV-PeV energy range. Events resulting from charged current (CC) interactions of muon (anti-)neutrinos will  have  a  track-like topology, while the CC interactions of the other neutrino flavors and neutral current (NC) interactions of all flavors will typically be observed as nearly point-like events. Hence,  the observed neutrino events are classified into two event classes: tracks and cascades.
\begin{figure}[h]
\begin{center}$
\begin{array}{cc}
\includegraphics[width=75mm]{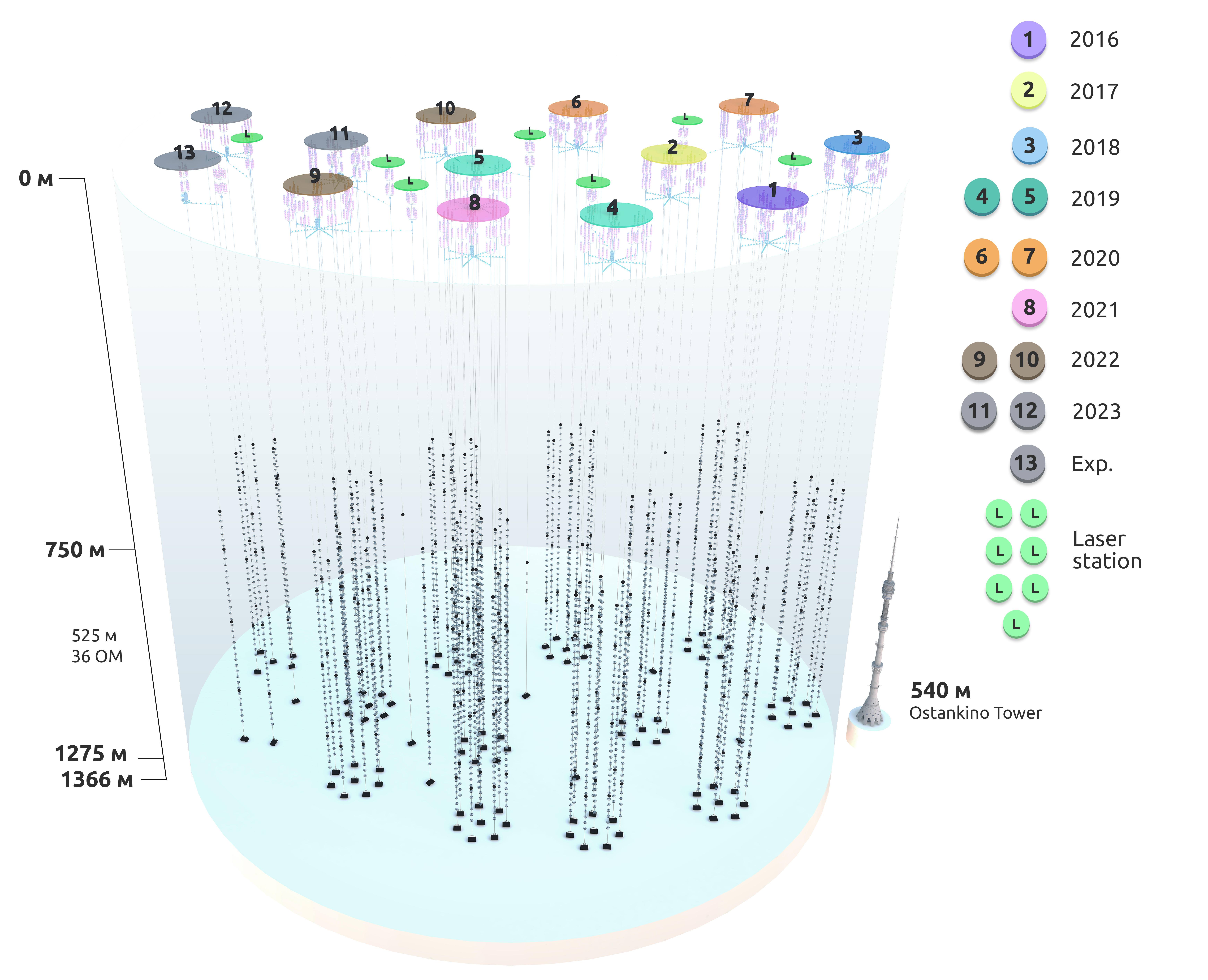}  &
\includegraphics[width=50mm]{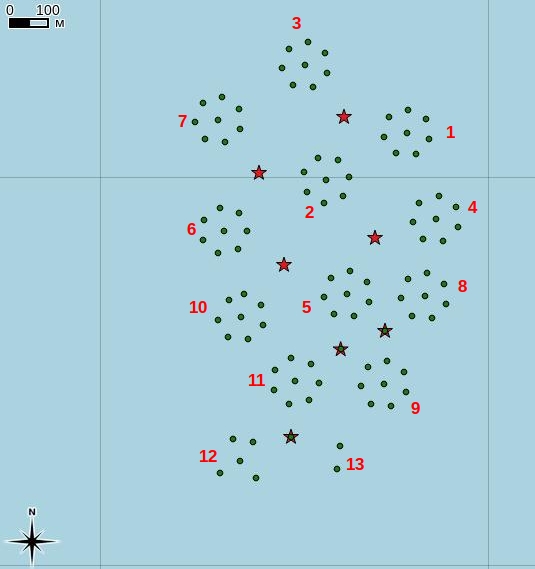}
\end{array}$
\end{center}
\caption{Left panel: Schematic view of the Baikal-GVD detector. The legend shows annual progress in the deployment. Right panel: Schematic view from the top of the Baikal-GVD detector for the year 2023. 
\label{fig.1}}
\end{figure}
The OMs are arranged on vertical load-carrying cables to form strings each anchored to the bottom of the lake and kept straight by a system of buoys at the top. Each string holds 36 optical modules. The optical module comprises a 10-inch  high-quantum-efficiency photo-multiplier tube (Hamamatsu  R7081-100 PMT), a  high  voltage unit and front-end electronics, all together enclosed in a pressure-resistant glass sphere. The OM is  also equipped with calibration LEDs and a set of digital sensors, including an accelerometer/tiltmeter, a compass, a pressure sensor, a humidity sensor, and two temperature sensors. The OMs are arranged with 15 m vertical spacing, for a total active string length of 525 m, starting 90 m above the lakebed. The PMT photocathodes are oriented downwards. A basic element of the detector readout in the Baikal-GVD is a section consisting of 12 OMs distributed vertically along the string, spaced by 15 m, and a control module (CM). Each OM of a section is connected to the CM by a 92 m long deep-sea cable. The CM controls the OM operation, converts the analog signals of the PMT into the digital form, forms local triggers of the section and time frames of events containing the pulse form. The conversion of analog signals is carried out by a 12-channel ADC with a sampling frequency of 200 MHz. The standard trigger condition requires a registration of two pulses on two neighboring OMs within the same section in a 100 ns time window with integrated charges exceeding channel-dependent thresholds. There are 3 sections installed on each string. The control of the sections operation and the exchange of the data is provided by a separate deep–sea electronic unit – string control module (SM). The string also holds acoustic beacons for acoustic monitoring of the PMT positions and LED beacons for the detector time calibration \cite{BaikalGVDPos,BaikalGVDCal}. Seven peripheral strings are uniformly located at a 60 m distance around a central one. Together they form a completely functionally independent unit called cluster, which is connected to the shore station via a dedicated electro-optical cable. Once the trigger condition is fulfilled for any of the CMs, a 5 $\mu$s event time frame is read out from all the CMs of the cluster. Each detector string can be recovered and re-deployed without the need to recover the whole cluster. The clusters are arranged on the lakebed in a hexagonal pattern, with a distance of 300 m approximately between the cluster centers.

The first cluster of the Baikal-GVD neutrino telescope was deployed in 2016 with 288 OMs. Since then, there has been an annual increase of the deployed clusters resulting in 3\,456 OMs attached to 96 vertical strings today. Timeline of the deployment within the years is shown in Fig.(\ref{fig.1}). Among the Baikal-GVD clusters there are special additional strings equipped with high-power pulsed lasers (red and black stars in Fig.(\ref{fig.1}) right panel) dedicated for the inter-cluster time calibration and the light propagation studies in Baikal water. 

According to a study made  with a specialized device, the light absorption length in the deep lake water reaches maximal values, $\sim$ 24 m, at a wavelength of 488 nm. The effective light scattering length is $\sim$ 480 m (at 475 nm). Both the absorption and scattering characteristics show minor variations over time (see \cite{ASP-15,RyabovICRC23} for details). Luminescent light of Baikal water is registered also by the OMs of the detector. Typically every year there are two periods of relatively stable optical background noise (OMs noise rates $\sim$ 40 kHz), which are intermitted by increased optical activity ($\sim$ 150 kHz during June-September period) \cite{LuminescenceICRC23}. In general, charge of these noise rate pulses is of a single p.e..

The most effective way to increase the Baikal-GVD telescope sensitivity of cascade-like neutrino events is to install additional inter-cluster strings (ICS) in the geometric centers of each triplet of the Baikal-GVD clusters. This is the result obtained by means of MC simulations performed for the configuration of 3 clusters with and without an ICS. For cascades with an energy more than 100 TeV, from the region where the background from atmospheric muons and neutrinos becomes smaller than the signal from astrophysical neutrinos, the increase in the number of events is 24$\%$ for the distance between clusters being 250 m. In case of muon events with a track length exceeding the geometrical dimensions of the telescope the effective area increased in proportion to the increased number of optical modules, i.e. there is no significant effect from the ICS installation.

The first inter-cluster string (ICS) was installed in 2022 and two more were commissioned in the expedition of 2023 (see Fig.(\ref{fig.1})). These external strings are situated approximately in the geometric centers of each three clusters of the detector. The design of the ICS is similar to a regular Baikal-GVD string. There are 36 OMs arranged in 3 sections. In each section, 12 OMs are connected to the control module (CM). Additional equipment that is atached to the ICS is connected to the CMs. These are acoustic modems (providing the positioning of the ICS) and a laser source with different levels of intensity and attenuation (providing inter-cluster time calibration and charge calibration of OMs). The ICS is installed as a ninth string of one of the clusters (for more details see \cite{AynutdinovICRC23}). 

\section{Progress in the track-like events analysis}

Muons produced in neutrino interactions manifest themselves in events of track-like topology in Baikal-GVD. Namely these are the events of charged current (CC) interactions of muon (anti-)neutrinos and tau (anti-)neutrino CC interactions in case when tau decays into a muon. Muon track reconstruction is a two stage process. 
\begin{figure}[h]
\begin{center}$
\begin{array}{cc}
\includegraphics[width=112mm]{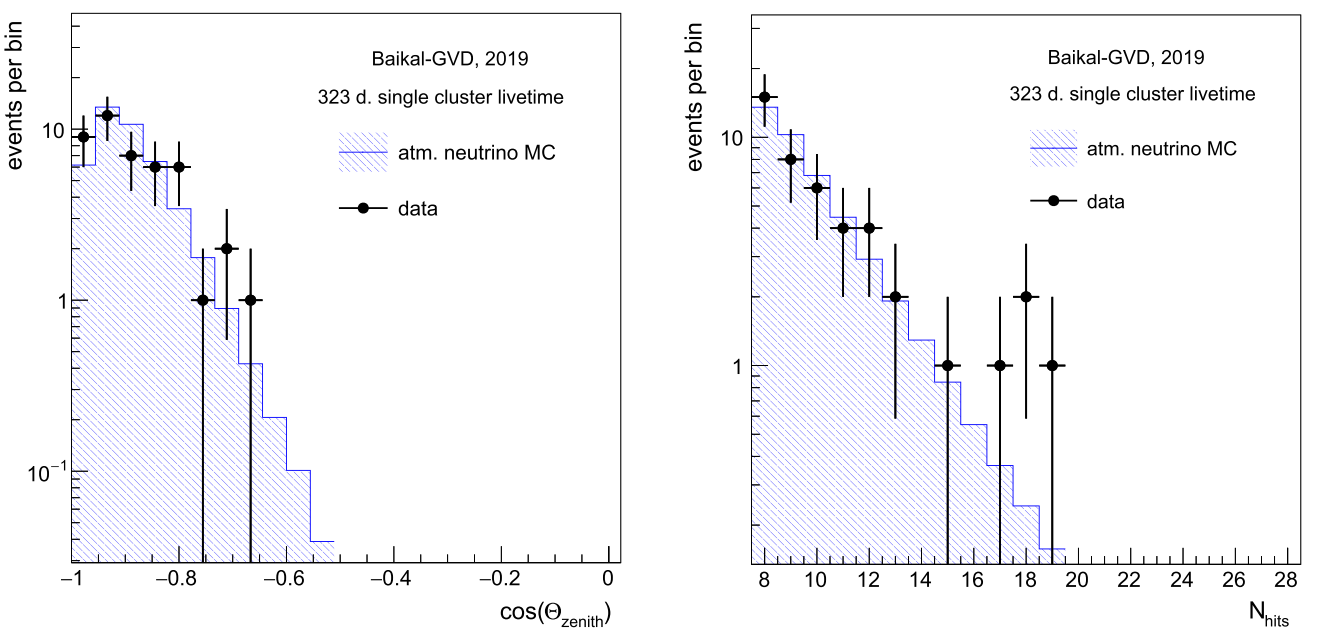}  & 
\includegraphics[width=25mm]{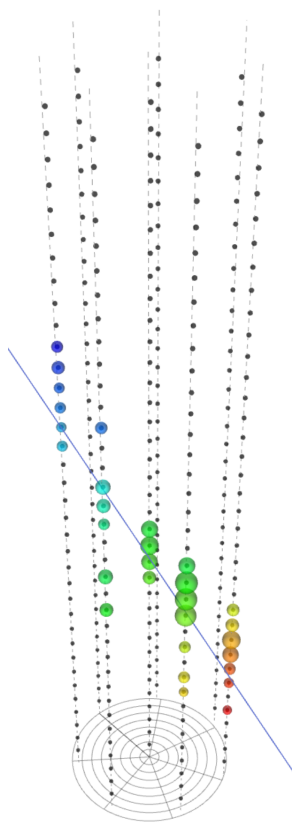}
\end{array}$
\end{center}
\caption{Upgoing track-like neutrino events as a result of the analysis from the 2019 dataset \cite{BaikalGVD-muon-alg}. Left panel: Zenith angle distribution. Middle panel: Distribution of number of hits used in the reconstruction. The black points are the Baikal-GVD data. Blue filled area stands for the MC prediction for atmospheric neutrinos. Right panel: Visualization of the 100 TeV upgoing muon neutrino candidate selected by the analysis. Estimated signalness > 90 $\%$ indicates with high probability that this event is of astrophysical origin. 
\label{fig.muon}}
\end{figure}

Firstly, the pulses generated by the Cherenkov light emitted by the muon are selected while the noise pulses from the luminiscense of the lake water or PMT dark current or afterpulses are excluded (see \cite{BaikalGVD-muon-ICRC23}). Secondly, the direction, energy, and various quality parameters of the track (e.g. fit convergence, value of minimisation function) are reconstructed.  Noise suppression for single-cluster reconstruction is performed with the effective algorithm based on the directional causality criterion and fast causally-connected pulse clique search algorithm \cite{BaikalGVD-muon-rec}. The precision of the track direction measurement depends on track length and varies from $\sim$ 1.5$^\circ$ for short tracks to below 0.5$^\circ$ for long nearly vertical tracks. 

The first set of 44 neutrino candidate events was selected from data collected in April–June 2019 by means of the fast single-cluster event reconstruction algorithm and a cut-based analysis \cite{BaikalGVD-muon-alg}. The rate of neutrino candidate events is in close agreement with the MC expectation for atmospheric neutrino flux (Fig.(\ref{fig.muon}) left and middle). The obtained event set is dominated by the atmospheric neutrinos with an average energy of 500 GeV. Reprocessing of the full year 2019 dataset allowed to find an upgoing muon with median estimate of energy corresponding to 103.4 TeV. The number of hits foung by the hit selection algorithm for this event is 30, the reconstructed zenith angle $\theta$ is 153.4$^{\circ}$ and visible track length is 332.4 m (see Fig.(\ref{fig.muon})). Considering the contributions from atmospheric muon bundles, atmospheric neutrino including prompt neutrino contribution, and using astrophysical neutrino spectrum index $\gamma_{astro}\sim$-2.36, the probability (signalness) of this event was estimated to be larger than 90$\%$.

\section{High-energy neutrino cascade events}

Reliable reconstruction of high-energy neutrino induced showers created in the vicinity of the sensitive volume of the detector is the key ingredient in the search for high-energy astrophysical neutrinos. In order to achieve this goal a set of cuts on quality variables were optimized on Monte Carlo simulations and tuned using the data sample accumulated in 2016 – 2017. Only pulses with charge Q higher than 1.5 p.e were considered. This condition allows to suppress background from water luminescence substantially. Another cut that helps to suppress the noise pulses is a criterium to select events with a large multiplicity of hit OMs $N_{hit}>7$ at least on three strings. 
\begin{figure}[h!]
\begin{center}$
\begin{array}{cc}
\includegraphics[width=65mm]{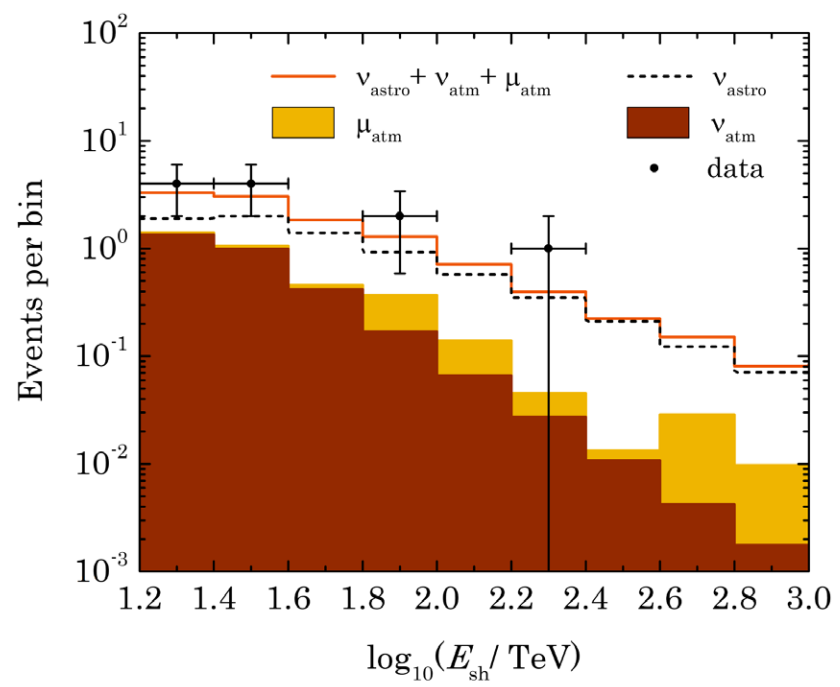}  &
\includegraphics[width=65mm]{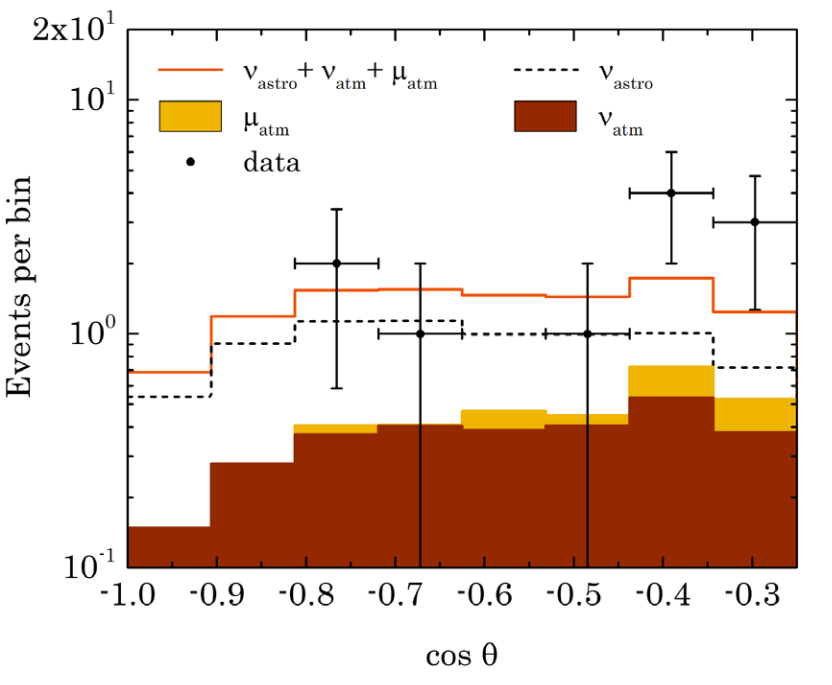}
\end{array}$
\end{center}
\caption{Left panel: Reconstructed cascade energy distributions of the dataset consisting of 11 selected events. The best-fit distribution of astrophysical neutrinos (dashed line), expected distributions from atmospheric muons (yellow) and atmospheric neutrinos (brown) and the sum of the expected signal and background distributions (orange line) are also shown. The atmospheric background histograms are stacked (filled colors). Right panel: The same for the reconstructed zenith angle distribution.
\label{fig.casc}}
\end{figure}

Reconstruction of energy,direction, and vertex coordinates of high-energy cascades is a two step procedure \cite{BaikalGVD-casc-rec}. The first step is to find the shower vertex coordinates $\vec{r}_{sh}$ by minimization of $\chi^2_t$ function with use of time information of pulses on OMs. The shower is assumed to be a pointlike source of light at this procedure level. The shower vertex coordinates are used as an input in the second step in which the shower energy ($E_{sh}$) and direction ($\theta$;$\phi$) are reconstructed by means of the maximum-likelihood method. Rejection of poorly reconstructed events is achieved by applying cuts on quality parameters, including the values of $\chi^2_t$ and maximum-likelihood function, OMs hit multiplicity $N_{hit}$. The precision of energy and direction of the shower crucially depends on the cascade energy, its position, and orientation relative to the cluster and varies typically between 10$\%$ – 30$\%$ and 2$^{\circ}$ – 4$^{\circ}$, respectively. 

For the search of astrophysical neutrinos data from 2018-2021 collected by the Baikal-GVD were used. Data analysis was performed by the procedure used in our previous studies \cite{BaikalGVD-cascades}. The outcomes are 11 high-energy cascade events with restrictions on OM hit multiplicity $N_{hit}>11$, reconstructed energy $E_{sh}>15$ TeV, and reconstructed zenith angle cos$\theta <$ -0.25 considered as astrophysical neutrino candidates. The energy and zenith angle distributions of these 11 events are shown in Fig.(\ref{fig.casc}) together with the distributions obtained by Monte Carlo simulation. This dataset of 11 detected events has been exploited to charaterize the diffuse astrophysical neutrino flux assuming single power law model of the form 
\begin{eqnarray}
  \Phi^{\nu+\bar{\nu}}_{astro}=3\times10^{-18} \Phi_{0} \left( \frac{E_{\nu}}{E_0} \right)^{-\gamma_{astro}} [\rm{GeV}^{-1}\rm{cm}^{-2}\rm{s}^{-1}\rm{sr}^{-1}]
\end{eqnarray}

\begin{figure}[h]
\begin{center}
\includegraphics[width=70mm]{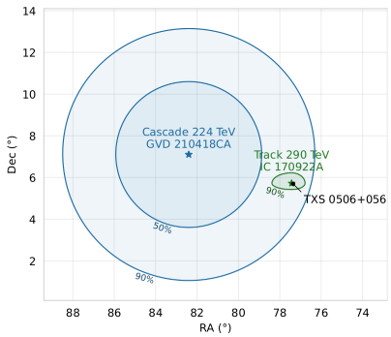}  
\end{center}
\caption{Equatorial coordinates of TXS 0506+056 blazar, high-energy (290 TeV) IceCube event IC 170922A with 90$\%$ confidence region, and the highest energy (224 TeV) upward-going Baikal-GVD event GVD 210418CA with 50$\%$ and 90$\%$ confidence region are shown.   
\label{fig.TXS56}}
\end{figure}
with $E_0 = 100$ TeV, one neutrino flavor flux normalization $\Phi_{0}$, and a spectral index $\gamma_{astro}$. The best fit parameters $\Phi_{0}=3.04$ and $\gamma_{astro}=2.58$ for the observed data are determined by a binned likelihood approach \cite{BaikalGVDflux}. The background-only hypothesis is excluded at the level of 3.05$\sigma$. It is worth to note that the Baikal-GVD results of cosmic neutrino diffuse flux measurements are consistent with measurements of IceCube and ANTARES (all-neutrino flavor).

The IceCube observatory has revealed that the strongest neutrino candidate source in extragalactic sky at E > 200 TeV is TXS 0506+056 \cite{IceCube2018}. We stress that the arrival direction of the highest energy upward-going neutrino candidate event in the Baikal-GVD data (GVD210418CA, E=224$\pm$75 TeV) is consistent with TXS 0506+056 blazar (see Fig.(\ref{fig.TXS56})). The signalness of this event is estimated to be 97.1$\%$. This neutrino source lies within a 90$\%$ confidence region of the GVD210418CA direction \cite{BaikalGVD-point-sources}.

\section{Conclusion}

The progress in the construction of the Baikal-GVD neutrino telescope that now comprises 3\,456 optical modules on 96 vertical strings has been reported. The rate of low-energy neutrino events observed in track-like event channel is in a good agreement with the MC expectation for atmospheric neutrino flux. First track-like event of high probability of astrophysical origin has been selected. The high-energy cascade event sample is consistent with the presence of astrophysical neutrino flux at the level of 3.05$\sigma$.

\end{document}